\documentclass[useAMS,usenatbib]{mn2e}

\usepackage{times}

\def\Om{\mbox{$\Omega_{\rm M}$}}
\def\OL{\mbox{$\Omega_{\Lambda}$}}
\def\Ox{\mbox{$\Omega_{\rm E}$}}
\def\Ok{\mbox{$\Omega_{k}$}}

\newcommand{\corr}{\mbox{$E_{\gamma}-E_{\rm peak}$}}

\def\spose#1{\hbox to 0pt{#1\hss}}
\newcommand\lsim{\mathrel{\spose{\lower 3pt\hbox{$\mathchar"218$}}
     \raise 2.0pt\hbox{$\mathchar"13C$}}}
\newcommand\gsim{\mathrel{\spose{\lower 3pt\hbox{$\mathchar"218$}}
     \raise 2.0pt\hbox{$\mathchar"13E$}}}

\title[A new method for using GRBs as cosmic rulers]
{A new method optimized to use Gamma Ray Bursts as cosmic rulers}
\author[Firmani et al.]{ Claudio Firmani$^{1,2}$\thanks{E-mail:
firmani@merate.mi.astro.it},
Gabriele Ghisellini$^{1}$, Giancarlo Ghirlanda$^{1}$ and Vladimir Avila-Reese$^{2}$\\
$^{1}$Osservatorio Astronomico di Brera, via E.Bianchi 46, I-23807 Merate, Italy\\
$^{2}$Instituto de Astronom\'{\i}a, U.N.A.M., A.P. 70-264, 04510, M\'exico, D.F., M\'exico}
\begin{document}


\pagerange{\pageref{firstpage}--\pageref{lastpage}} \pubyear{2002}

\maketitle

\label{firstpage}

\begin{abstract}
We present a new method aimed to handle long Gamma--Ray Burst (GRBs) as cosmic rulers. 
The recent discovery of a tight correlation between the collimation corrected GRB 
energy and the peak of the $\gamma$--ray spectrum has opened the possibility to use 
GRBs as a new category of standard candles. 
Unfortunately, because of the lack of low--$z$ GRBs, up to now this correlation is 
obtained from high--$z$ GRBs with the consequence that it depends on the cosmological 
parameters we pretend to constrain. 
Hopefully this circularity problem will be solved when, in a few years, the low--$z$ 
GRB sample will be increased enough.
In the meanwhile we present here a new Bayesian method that eases the aforesaid 
circularity problem, and allows to introduce new constrains on the cosmological 
$(\Om,\OL)$ diagram as well as to explore the universe kinematics up to $z \approx 3$.
The method we propose offers the further advantage to make handy the problem of the
$(\Om,\OL)$ loitering line singularity which inevitably appears when standard
candles with $z>2$ are used.
The combination of GRB with SN Ia data makes the popular $\Lambda$CDM cosmology 
more consistent with the Hubble diagram at a 68\% confidence level. 
For a flat cosmology we find $\Om=0.28\pm0.03$ for the combined GRB+SN Ia data set.  
Correspondingly, the transition redshift between cosmic deceleration and acceleration 
is $z_{\rm T}=0.73\pm0.09$, slightly larger than the value found by considering SNe Ia 
alone.  
We briefly discuss our results also in terms of non--$\Lambda$CDM dark energy models. 

\end{abstract}

\begin{keywords}
cosmological parameters  --- cosmology:observations --- distance 
scale---gamma rays: bursts
\end{keywords}

\section{Introduction}
\begin{figure*}
\begin{center}
\vspace{8cm}
\includegraphics{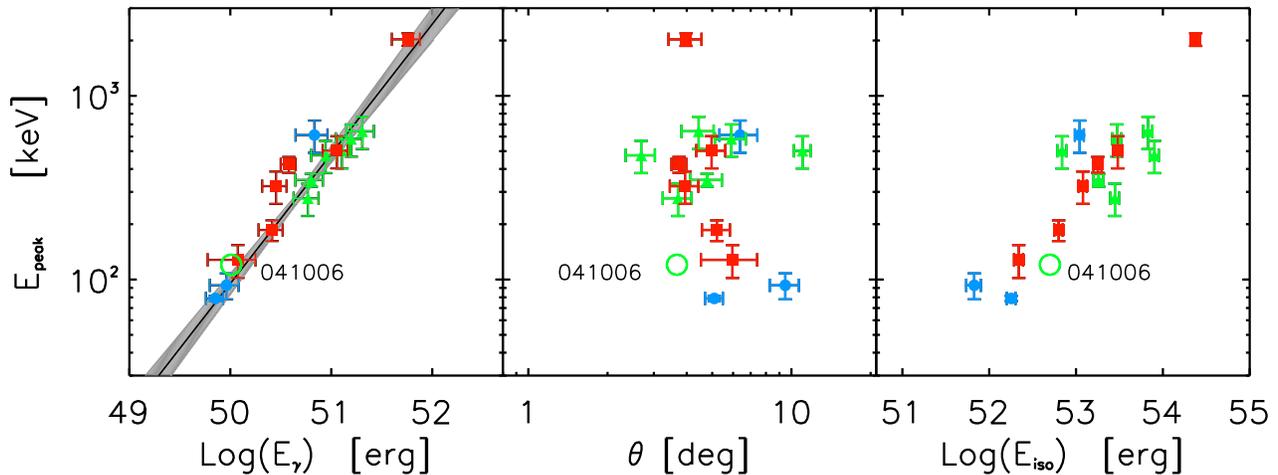}
\vskip -0.8 true cm
\caption{  
Rest frame  correlations between the spectral peak energy
$E_{\rm peak}$  and the collimation corrected energy $E_\gamma$ (left
panel),  the collimation  angle  $\theta$ (middle)  and the  isotropic
bolometric energy (right), for the 15 GRBs considered in GGL, assuming
\Om=0.3 and \OL=0.7.  Different  symbols  (colors)  correspond  to
different  redshift  ranges:  circles  (blue), triangles  (green)  and
squares   (red)   are   for   $z<0.7$,  $0.7<z<1.55$   and   $z>1.55$,
respectively.  The  solid line in  the left panel represents  the best
fit $E_{\rm peak}$--$E_{\gamma}$ correlation ($\chi^{2}=1.27$) and the
shaded region  its uncertainty computed in the  coordinate frame where
the variances on its parameters are uncorrelated. 
We also show (open circle) the recent GRB 041006
(data in Galassi et al. 2004, D'Avanzo et al. 2004 
and the HETE2 web page
(http://space.mit.edu/HETE/Bursts/GRB041006).
}
\vskip -0.5 true cm
\end{center}
\end{figure*}

The Hubble diagram of type Ia Supernovae (SNe Ia) has revolutionized
cosmology, showing that the current expansion of the universe is
accelerated (Riess et al. 1998; Riess et al.  2004; Perlmutter et al.
1999), with a possible transition from deceleration to acceleration at
low redshifts (Riess et al. 2004; Dicus \& Repko 2004).  
The most popular explanation for the recent acceleration implies the 
existence of a dominant energy density component with negative
pressure (dark energy, see for a recent review Sahni 2004).  
Its physical nature is still a mystery.  The simplest
candidate for dark energy is the modern version of the Einstein
cosmological constant $\Lambda$, which, however, raises several
theoretical difficulties (see Padmanabhan 2003; Peebles \& Ratra 2003;
Sahni 2004). There is now a flurry of activity focused to unveil the
nature of dark energy or to constrain any alternative explanation to
accelerated expansion. The forthcoming SNAP mission
\footnote{http://snap.lbl.gov/} ($z_{\rm max}\approx 1.7$) will offer
an unprecedented accuracy on this undertaken. 
Besides accuracy, also a survey reaching larger redshifts is desirable in
order to overcome parameter degeneracies, and to study the evolution
of dark energy (Linder \& Huterer 2003; Weller \& Albrecht 2002;
Nesseris \& Perivolaropoulos 2004).  Furthermore,
alternative
standard candles to SNe Ia, free of extinction issues, are also highly
desired.  Long GRBs have been pointed out as possible candidates to be
these high--redshift candles (Schaefer 2003), and in Ghirlanda et
al. (2004b, GGLF thereafter) we have shown that they are indeed.  

Assuming isotropy, the
energy emitted by GRBs in $\gamma$--rays ($E_{\rm iso}$) is spread over 4
orders of magnitude, but the presence of an achromatic break in their
afterglow lightcurve is a strong evidence that their emission is
collimated into a cone of semiaperture angle $\theta$ (e.g. Rhoads
1997; Sari, Piran \& Halpern 1999).  Correcting for this anisotropy,
the resulting $\gamma$--ray energetics
$E_{\gamma}=(1-\cos\theta)E_{\rm iso}$ clusters around $E_\gamma\sim
10^{51}$ erg, with a small dispersion (0.5 dex, Bloom, Frail \&
Kulkarni 2003; Frail et al. 2001), yet not small enough for a
cosmological use (Bloom, Frail \& Kulkarni 2003).

Recently, a very tight correlation has been found (Ghirlanda,
Ghisellini \& Lazzati 2004a, GGL thereafter) between $E_{\gamma}$ and
the peak energy $E_{\rm peak}$ of the prompt emission (in a $\nu
F_\nu$ representation) (\S 2).  This correlation has been used for a
reliable estimate of $E_{\rm iso}$, and therefore of the luminosity
distance, with the aim to construct a Hubble diagram up to $z=3.2$
(GGLF; Dai, Liang \& Xu 2004).

The existence of the $E_{\gamma}$--$E_{\rm peak}$ correlation,
allowing to know the intrinsic emission of GRBs, makes them {\it
standard candles} similar to SNe Ia and Cepheid stars.  Hopefully,
this correlation will be calibrated from low--$z$ GRBs, and
corroborated by a robust theoretical interpretation.  Currently
low--$z$ GRBs are scarce: we must wait for a few years before having a
sufficient number of GRBs at $z<$0.2.  For the time being, the
correlation is obtained from high--$z$ GRBs with the consequence that
it depends on the cosmological parameters {\it which we pretend to
find}.  We present here a new method with respect to GGLF that eases
this difficulty making possible the full use of the available
information.  We will assume that the $\Lambda$CDM cosmology has
$h=0.71$, $\Om=0.27$, and $\OL=0.73$.

\section{The $E_{\gamma}$--$E_{\rm peak}$ correlation}

In Fig. 1 (left panel) we show the correlation between the rest frame
$E_{\rm peak}$ and $E_\gamma$ for the 15 bursts considered by GGL,
using the data listed in their Tab. 1 and Tab.  2, and assuming
$\Omega_{\rm M}=0.3$ and $\Omega_{\Lambda}=0.7$.  With this choice of
cosmology, the reduced $\chi^2=1.27$ for the fit with a powerlaw.  The
mid panel shows $E_{\rm peak}$ as a function of the jet aperture angle
$\theta$, while the right panel shows $E_{\rm peak}$ as a function of
the isotropic emitted energy $E_{\rm iso}$ (the ``Amati" relation,
Amati et al. 2002) for the same 15 bursts. 
Also shown is the recent GRB 041006:
although fully consistent with the GGL correlation, it
is not included in the sample of 15 GRBs used for cosmology
due to the still unpublished uncertainties on the
relevant parameters.
The shaded area in the
left panel shows the region of uncertainties of the best fit
correlation:
$E_{\rm peak}/(263\pm 15 ~{\rm keV}) = 
[E_\gamma / (4.2\times 10^{50} {\rm erg}) ]^{0.706\pm0.047}$.
The errors on its slope and normalization are calculated in the
``baricenter" of $E_{\rm peak}$ and $E_{\gamma}$, where the slope and
normalization errors are uncorrelated (Press et al. 1999).  For similar
$E_{\rm peak}$, there is a distribution of jet aperture angles, which
is why $E_\gamma$, for a given $E_{\rm peak}$, is much less spread
than the corresponding $E_{\rm iso}$ values.

In the present sample of bursts with known redshift GRB~980425
 (Galama et al.  1998) and GRB~031203 (Gotz et al. 2003) are outliers
 for both the $E_{\rm peak}-E_{\rm iso}$ and the \corr ~correlation.
 These events are indeed {\it peculiar} for their small prompt and
 afterglow emission, as recently discussed by Soderberg et
 al. (2004a).  It has been also suggested that these two events are
 sub--energetic, with respect to ´´classical´´ GRBs, because they are
 observed off--axis (Waxman 2004; Yamazaki, Yonetoku \& Nakamura 2003).
 Also GRB 020903 (Soderberg et al. 2004b) may be an outlier for the \corr
 ~correlation, although, as discussed in GGL, its afterglow data are
 not conclusive.

\section{The method}

A Friedmann-Robertson-Walker cosmological model is defined by the
Hubble parameter $H(z)$ given by: 
\begin{equation}
H^2(z) = H_0^2 \left[\Om (1+z)^3 + \Ok (1+z)^2 +\Ox f(z)\right] 
\end{equation}
where $\Ok =1-\Om-\Ox,$ $\Om$ and $\Ox$ are the
present--day density parameters of matter and dark energy components,
and $f(z)$ is a function that depends on the equation-of-state index
of the dark energy component, $w(z)=p_{\rm E}(z)/\rho_{\rm E}(z)c^2$.
Later on we will take into account the case of a cosmological constant
($\Lambda$--models) with $\Ox=\OL$, $w=-1$ and $f(z)=1$, as well as a
linear parametrization of the dark energy equation of state
($W_i$--models) adopting $w(z)=w_0+w^\prime z$, a flat cosmology, and
$f(z)=(1+z)^{3(1+w_0-w^\prime)}\exp(3w^\prime z)$.  
Once defined $H(z)$, the luminosity distance as a function of $z$ 
(Hubble diagram) is calculated straightforwardly.  
The transition redshift $z_{\rm T}$ from deceleration to acceleration 
for $\Lambda$--models is given by $2 \OL = \Om (1+z_{\rm T})^3$, 
and for W$_i$--models by 
$\Om =(1-\Om )(-1-3w_0-3w^\prime z_{\rm T}) (1+z_{\rm T})^{3(w_0-w^\prime )}
\exp(3w^\prime z_{\rm T})$.  
In this letter we shall fit observations with only two free parameters.  
These ones are
(i) $(\Om,\OL)$ for $\Lambda$--models, representative of the cosmological
constant models, 
(ii) $(\Om,w_0)$ with a flat cosmology and $w^\prime=0$ for $W_0$--models, 
which identify quiessence dark energy models, and finally 
(iii) $(w_0, w^\prime)$ with $(\Om=0.27, \OL=0.73)$ for $W_1$--models, 
to identify kinessence models.

The use of GRBs as cosmic rulers brings up two major difficulties:

1) Due to the current poor information on low--$z$ GRBs, the \corr
~correlation (and its scatter) depends on the assumed cosmology. This
difficulty hides a circularity problem and has to be handled with care 
otherwise it leads to a tautology: if a cosmology is used to fix the 
\corr~ correlation then the \corr~ correlation cannot be used to find 
a cosmology. 
This problem was not taken into account by Dai, Liang \& Xu (2004).

2) Because of the previous point and the complex behavior of the
luminosity distance $d_{L}$ as a function of $z$, \Om, \Ox, some mathematical 
undesirable attractors appear.  
This effect can be seen in Fig. 1 of GGLF where the GRB confidence level 
curves are tangent near \Om=0.07 and \OL=1.15.  
The effect is basically due to the rather singular behavior 
of $d_{L}$ near the loitering line that discriminates between Big--Bang
and no Big--Bang universes (close to the region probed by GRBs, but
not by SNe).  Here $d_{L}$ falls to zero for $z\ga$1.  The sharpness
of this fall increases with $z$.  The obvious result is that for high
$z$ the constant $d_{L}$ lines tend to be tangent to the loitering
line converging towards \Om$\sim$0 and \OL$\sim$1.  Due to the high GRB
redshifts, this is just the region where GRBs lie (this problem is
practically irrelevant for SNe Ia).  If the statistical approach is not
sufficiently powerful (as the $\chi^2$ method used in GGLF), the
singularity of $d_{L}$ is able to attract the most probable solution
to some improper value (for the $\chi^2$ method used in GGLF this
solution is, in fact, \Om$\sim$0.1 and \OL$\sim$1.1).  

It is necessary to introduce a new powerful statistical approach 
in order to avoid mathematical artificial tricks in the GRB case.
{\it The Bayesan approach we propose below solves the difficulties
stated before} and it might come in useful in other similar problems. 
 
In the case of $\Lambda$--models, for any given cosmology
$\bar{\Omega}\equiv (\bar{\Omega}_M,  \bar{\Omega}_\Lambda)$ 
there is a ``best  fitted"  $E_\gamma$--$E_{\rm  peak}$  correlation. 
For each GRB this correlation allows to calculate an {\it observed} 
$d_{Li}(\bar{\Omega}) = [ E_{\rm iso} (1+z)/(4\pi F)]^{1/2}$,
where $F$ is the fluence and $E_{\rm iso}$ is derived following
Tab. 1, Tab. 2  and Eq. 1 of GGL.
Now, on the Hubble diagram, for each cosmology
$\Omega$ we can define 
$\chi^2(\Omega,\bar{\Omega}) = \sum_{i=1}^N(d_{Li}(\bar{\Omega})-
d_L(z_i,\Omega))^2/\sigma_i^2$,
where $d_L(z_i,\Omega)$ is the {\it theoretical} luminosity distance, 
$\sigma_i$ the standard deviation on $d_{Li}(\bar{\Omega})$, 
and the sum is extended on the $N$ GRBs of the sample.
Making use of the incomplete gamma function (Press et al. 1999), we 
transform $\chi^2(\Omega,\bar{\Omega})$ into the conditional probability
$P(\Omega|\bar\Omega)$ that provides  the probability for each $\Omega$ 
given a possible $\bar\Omega$--defined correlation.
In order to understand the essential of the  method, we can imagine that 
at first all $\bar\Omega$s (i.e., the corresponding correlations) are assumed 
to be equally probable. Thanks to the previous conditional probability, it follows 
that  we do have some knowledge of which is the probability of $\bar\Omega$.   
This is derived from the same $P(\Omega|\bar\Omega)$ adding the contributions 
of all possible $\bar\Omega$--defined correlations.
Taking this probability we assign now a weight to $\bar\Omega$, then we repeat 
the cycle.  
We arrive to another set of probabilities for $\Omega$.  
We continue to iterate, until convergence.
This intuitive approach find a simple and elegant formalization through a
Bayesian approach by the formula
\begin{equation}
P(\Omega) = \int P(\Omega|\bar\Omega)P^\prime(\bar\Omega) d\bar\Omega
\end{equation}
where  the integration is extended on the accessible $\Omega$ plane.  
The {\it prior  probability} $P^\prime$  defines the  weights for  the  
$E_\gamma$--$E_{\rm peak}$ correlation, 
the {\it likelihood} $P(\Omega|\bar\Omega)$ is the conditional probability
defined above, 
and  $P$ is  the {\it  posterior  probability} that measures the goodness 
of fit of the possible cosmologies.
We convert the  previous formula into an equation setting $P^\prime=P$.
A Montecarlo method provides an economical approach to find the numerical 
solution of this equation.   
For $W_0$--models the previous procedure is applied to the parameter pair
$(\Om,w_0)$, and for $W_1$--models the same is applied to $(w_0,w^\prime)$.

This approach is quite different from mapping the $\chi^2$ parameter
for all points in the $\Omega$ plane by minimizing the scatter of the
data points around a $\Omega$-dependent correlation (GGLF, see below).

\begin{figure}
\vspace{9cm}
\includegraphics{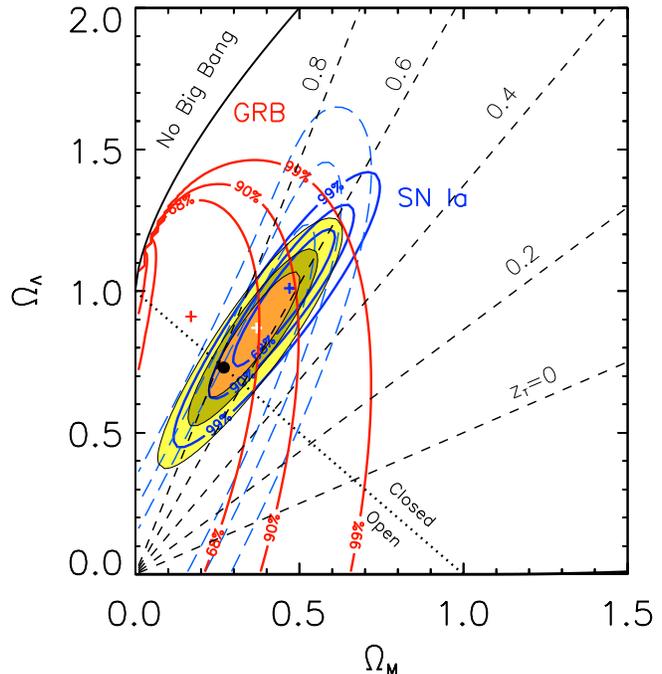}
\caption{ 
Constraints in the \Om--\OL ~plane derived from our GRB
sample (15 objects, red contours), from the ``Gold" SN Ia sample (156
objects; blue solid lines, derived assuming a fixed value of $H_0=65$
km s$^{-1}$ Mpc$^{-1}$, making these contours slightly different from
Fig. 8 of (Riess et al. 2004), and from the subset of SNe Ia at $z>0.9$
(14 objects, blue long dashed lines).  The three colored ellipses are
the confidence regions (orange: 68\%; light green: 90\%; yellow: 99\%)
for the combined fit of type Ia SN+GRB samples.  Dashed lines
correspond to the changing sign of the cosmic acceleration
[i.e. $q(z)=0$] at different redshifts, as labelled.  Crosses are the
centers of the corresponding contours (red: GRBs; blue: SNe Ia, white:
GRB+SN Ia).  The black dot marks the $\Lambda$CDM cosmology.  The
dotted line corresponds to the statefinder $r=1$, in this case it
coincides with the flatness condition.}
\end{figure}

\section{Results}

For fitting the cosmological parameters we have selected the 15 GRBs reported
on GGL (see  \S 2).  The GRB sample extends  from $z=0.16$ to $z=3.2$.
The SN  Ia sample, that we  combine with the GRB  sample, includes the
most  recent  discovered high  redshift  SNe Ia  (up  to  1.7) and  is
composed by 156 objects (the ``Gold'' sample, Riess et al. 2004).

For the $\Lambda$--model, we constrain $\Om$ and $\OL$.  Fig. 2 shows
the confidence levels (CL) for (i) GRBs alone (adopting our Bayesian 
method), (ii) SNe Ia alone (for the entire ``Gold" sample and for only
the subset of 14 SNe Ia at $z>0.9$, using in both cases the standard
approach), and (iii) the combination of all SNe Ia and GRBs.  The
dotted line shows the statefinder parameter $r=1$ (Sahni et al. 2003;
see also the jerk parameter in Visser 2004, and Riess et al. 2004)
that in $\Lambda$--model coincides with the flatness condition.

The difference between Fig. 2 and Fig. 1 of GGLF is remarkable. In
Fig. 2 the most probable solution is shown by the red cross which
corresponds to \Om=0.17 and \OL=0.91, while in Fig. 1 of GGLF the best
fit model corresponds to \Om$\simeq$0.1 and \OL$\simeq$1.1.

For GRBs the CL is a little larger than for high--$z$ SNe Ia, 
but this CL is expected to be reduced by new experiments. 
Note that the orientations of the CL
contours are related to the average redshift of the sample; 
the increase of this redshift from all SNe Ia
to high--$z$ SNe Ia alone, and from high--$z$ SNe Ia to GRBs produces a 
counterclockwise rotation of the CL contour pattern.

\begin{figure}
\vspace{9cm}
\includegraphics{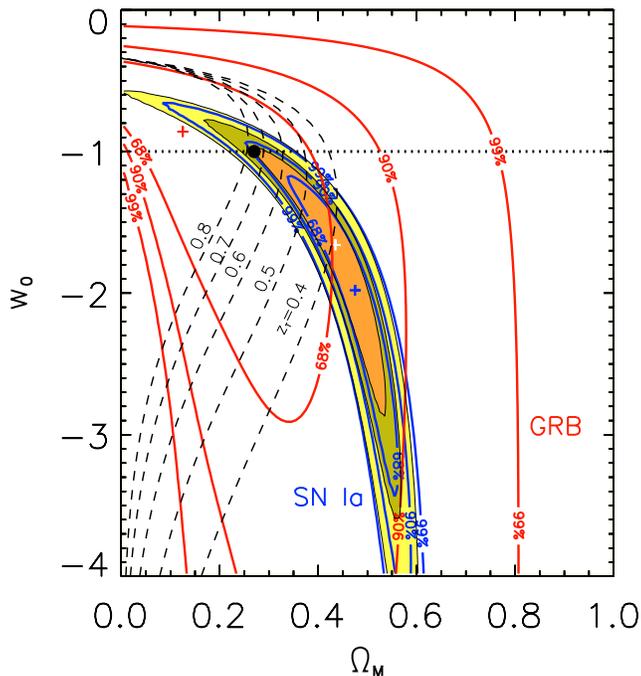}
\caption{
Constraints on $w_0$ and \Om\ for a
flat cosmology with dark energy whose equation of state is constant  
($w_0=-1$ corresponds to a cosmological constant). 
To construct the GRB contours we have followed the same method as in
Fig. 2.  
Colored regions: combined (GRB+SN Ia) constraints (color code as in Fig. 2).
Dashed lines correspond to the changing sign of the cosmic acceleration 
[i.e. $q(z)=0$] at different redshifts, as labelled.  
Crosses are the centers of the SNe Ia contours (blue: SNe Ia
alone, white: GRB+SN Ia). 
The black dot marks the $\Lambda$CDM cosmology.
The dotted line corresponds to the statefinder $r=1$.}
\end{figure}

When we combine SNe Ia with GRBs the best--fit model moves from
$\Om=0.48$, $\OL=1.00$ for the SN Ia sample to $\Om=0.37$, $\OL=0.87$
for the combined SN+GRB sample, favouring a closed universe.  However,
due to the high correlation of the CL contours in Fig. 2, the
best--fit value does not mean much.  Instead, it is worth to remark
that the flat geometry solution $\Om+\OL=1$ is compatible now with the
68$\%$ CL.  Assuming a flat geometry prior we obtain
$\Om=0.28\pm0.03$, in excellent agreement with independent galaxy
clustering measurements (Hawkins et al. 2003; Schuecker et al. 2003)
and the $\Lambda$CDM cosmology.  In Fig. 2 we also plot the curves of
constant $z_{\rm T}$.  For our best--fit model, $z_{\rm T}=0.67$,
while for the flat geometry case our constraints give 
$z_{\rm T}=0.73\pm 0.09$.

For $W_0$--models Fig.  3 shows the constraints on $w_0$ vs $\Om$
(i.e.  a constant equation of state).  Best values for the SN+GRB
sample are $\Om =0.44$ and $w_0=-1.68$ with $z_{\rm T}=0.40$, while
the $\Lambda$CDM is compatible with the 68$\%$ CL (it was not for SN
Ia alone).  Inside the uncertainty, $w_0<-1$ ($r>1$) quiessence dark
energy models appear favoured.  However, while SN Ia best--fit model
has $w_0<-1$, GRBs alone prefer $w_0>-1$, even if close to $-1$.  A
more complete sample of both objects is expected to support
$\Lambda$CDM.  For $W_1$--models Fig.  4 shows the constraints on the
dark energy parameters $w_0$ vs $w^\prime$.  For the SN+GBR sample the
best--fit values are $w_0=-1.19$ and $w^\prime=0.98$ with 
$z_{\rm T}=0.55$.  The $\Lambda$CDM again is now fully within the 68$\%$ CL of
the combined SN+GRB fit.  Similar to the previous case even now,
inside the uncertainty, $r>1$ kinessence models (i.e.  Chaplygin gas,
type 1 braneworld models, see Alam et al. 2003) seem favoured.  In
both $W_{i}$, SN Ia alone marginally reject $\Lambda$CDM
cosmology. This tendency is balanced by GRBs making hopeful a
$\Lambda$CDM cosmology for the combined sample. {\it We have here a
clear example how GRBs complement the cosmological information derived
by SN Ia.}

\begin{figure}
\vspace{9cm}
\includegraphics{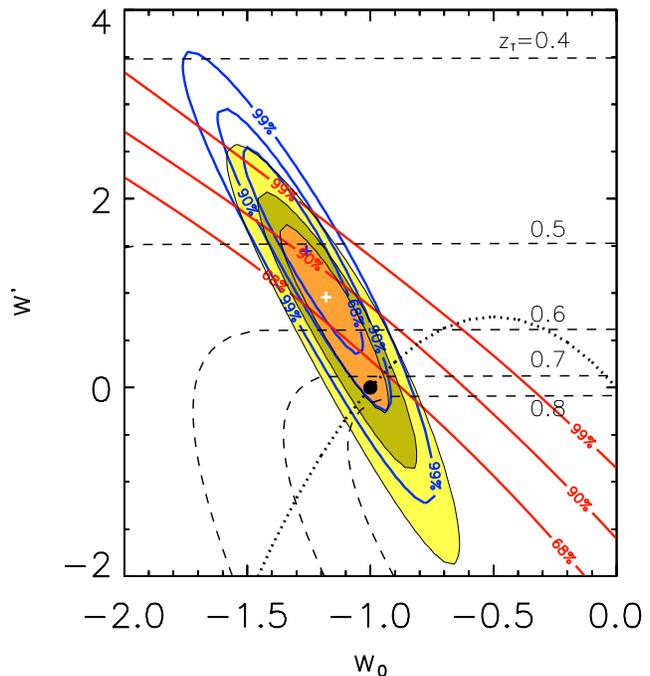}
\caption{Constraints on the $w_0$ and
$w^\prime$ parameters entering the dark energy equation of state 
(see \S 3; $w_0=-1$ and $w^\prime=0$ correspond to the cosmological
constant \OL).  We assume a flat geometry and \Om=0.27.  Blue contours: 
constraints from type Ia SNe; red contours: constraints from our GRBs.  
To construct the GRB contours we have followed the same method as in
Fig. 2.  Colored regions: combined (GRB+SN Ia) constraints (color code
as in Fig. 2).  
Dashed lines correspond to the changing sign of the cosmic acceleration 
[i.e. $q(z)=0$] at different redshifts, as labelled.  
Crosses are the centers of the SNe Ia contours (blue: SNe Ia
alone, white: GRB+SN Ia). 
The black dot marks the $\Lambda$CDM cosmology.
Dotted line corresponds to the statefinder $r=1$.}
\end{figure}

\section{Discussion}

Several kinematical and dynamical parametrizations of cosmology, 
aimed to optimize the fittings to the available observations, have 
been proposed. 
However current Hubble diagrams of SNe Ia need to be extended to 
higher redshifts to constrain such parameters with better accuracy.

GRBs are the natural objects able to extend the measure of the universe 
to very high redshifts. This is now possible thanks to the $E_{\rm peak}$ vs.
$E_\gamma$ empirical correlation, that allows to estimate $E_{\rm iso}$
of each GRB.  At present this correlation is known on the ground of 15 
GRBs distributed on redshifts up to 3.2; therefore its knowledge
depends on the assumed cosmology. We presented here a new method based on
a Bayesian approach that takes into account our lack of 
knowledge on the correlation, and optimizes the information to derive
the best--fit model and the CL contours.

Despite the small number of useful GRB events and their somewhat large
observational errors, a clear trend emerges when combining GRB and SNe
Ia data.  The inclusion of GRBs makes the $\Lambda$CDM cosmology
compatible with the 68$\%$ CL.  A similar trend, but less pronounced,
is observed for the SN Ia sample alone when including and not the 16
HST high--$z$ SNe (see also Choudhury \& Padmanabhan 2004; Alam et
al. 2004).

Fig. 5 shows the Hubble diagram in the form of residuals with
respect to the specific choice of \Om=0.27 ~and \OL=0.73, for SNe Ia
and GRBs, together with different lines corresponding to different
\Om, \OL ~pairs. It is likely that the GRB event rate follows or
increases with $z$ faster than the global star formation rate (e.g.,
Lloyd--Ronning, Fryer \& Ramirez--Ruiz 2002; Yonetoku et al. 2003;
Firmani et al. 2004), ensuring that GRBs exist up to
$z\sim$10--20. Even at these redshifts, GRBs are easily detectable,
but the ability of high--$z$ GRBs to accurately measure the Universe
depends of course on the errors of the relevant observables.  The fact
that the bolometric fluences of GRBs with measured $z$ do not
strongly correlate with $z$ (GGL) means that high--$z$ bursts can be
observed with large signal-to-noise ratios.  This ensures that 
$E_{\rm iso}$ and $E_{\rm peak}$ can be measured with good accuracy also at
high $z$'s. To derive $E_\gamma$, we need also information related to
the afterglow emission.  This, for high--$z$ GRBs observed at a given
time $t$ after trigger, is related to the intrinsic emission at an
earlier $t/(1+z)$ time, when the emitted flux was stronger.  It
follows that the observed afterglow fluxes of high--$z$ GRBs are not
much fainter than the afterglow fluxes of nearby GRBs observed at the
same time (Lamb \& Reichart 2000).  Thus, high-- and low--$z$ GRBs
will have comparable error bars on the relevant quantities required to
derive their luminosity distance.
In addition, the detection of GRBs is not limited nor affected by 
dust extinction or by optical absorption by Ly--$\alpha$ clouds, 
which are instead important issues for high--$z$ SNe Ia.

\begin{figure}
\vspace{5.2cm}
\includegraphics{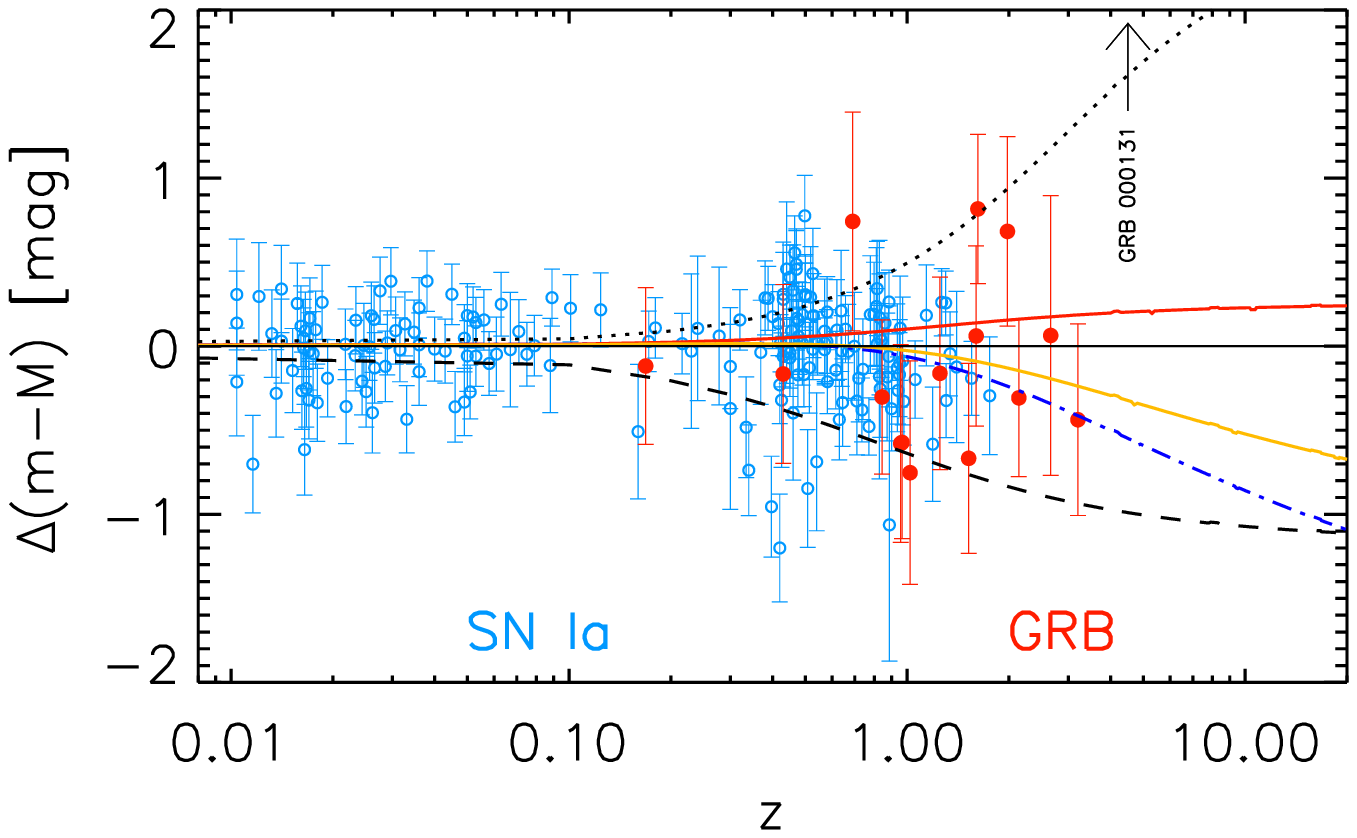}
\caption{
Residuals of the distance moduli of
SNe Ia and of our GRBs with respect to the $\Lambda$CDM 
case \Om=0.27, \OL=0.73.  
Also shown are the differences of various
other cosmological models with respect to the \Om=0.27, \OL=0.73
~case: 
red solid line: \Om=0.2, \OL=0.8; 
orange solid line: \Om=0.37, \OL=0.85; 
blue dot--dashed line: \Om=0.45, \OL=0.95; 
black dashed line: \Om=1, \OL=0; 
black dotted line: \Om=0.01, \OL=0.99. 
The arrow
marks the redshift of GRB 000131 ($z$=4.5) which is the most distant
GRB of known $z$.  Since for this GRB there is
no measured jet--break time, it is not included in our sample.}
\vskip -0.5 true cm
\end{figure}


\section*{Acknowledgments}
We thank Davide Lazzati for useful discussions and Giuseppe Malaspina
for technical support.
We thank the italian MIUR for funding through Cofin 
grant 2003020775\_002.


\begin{thebibliography}{99}

\bibitem[Alam, Sahni, \& Starobinsky(2003)]{alam03}
Alam U., Sahni V., Saini T.D. \& Starobinsky A.A. 2003, preprint (astro-ph/0303009)

\bibitem[Alam, Sahni, \& Starobinsky(2004)]{alam04}
Alam U., Sahni V. \& Starobinsky A.A. 2004, preprint (astro-ph/0403687)

\bibitem[Amati et al.(2002)] {amati02}
Amati L. et al. 2002, A\&A,  390, 81


\bibitem[Bloom, Frail \& Kulkarni(2003)]{bloom03}  
Bloom J.~S., Frail D.~A. \& Kulkarni S.~R. 2003, 
ApJ, 594, 674

\bibitem[Choudhury \& Padmanabhan(2004)]{CP04}
Choudhury T.R. \& Padmanabhan T. 2004, preprint (astro-ph/0311622)

\bibitem[Dai, Liang \& Xu(2004)]{dai04} 
Dai Z.G., Liang E.W. \& Xu D. 2004, ApJ, 612, L101

\bibitem[davanzo]{davanzo04}
D'Avanzo P., et al., 2004, GCN 2788

\bibitem[Dicus \& Repko(2004)]{dicus04}
Dicus D.A \& Repko W.W. 2004, preprint (astro-ph/0407094)


\bibitem[Firmani et al.(2004)]{firmani04} 
Firmani C., Avila--Reese V., Ghisellini G. \& Tutukov A.V. 2004,
ApJ, 611, 1033

\bibitem[Frail et al.(2001)]{frail01}
Frail D.~A. et al. 2001, ApJ, 562, L55

\bibitem[Galama et al. (1998)]{gala98}
Galama T. et al. 1998, Nature 395, 670

\bibitem[Galassi]{galassi04}
Galassi M. et al., GCN 2770


\bibitem[Ghirlanda et al.(2004a)]{ggl04} 
Ghirlanda G., Ghisellini G. \& Lazzati D.
2004a, ApJ, 616, 331 (GGL)

\bibitem[Ghirlanda et al.(2004b)]{gglf04}
Ghirlanda G., Ghisellini G., Lazzati D. \& Firmani, C. 2004b,
ApJ, 613, L13 (GGLF)

\bibitem[Gotz et al. (2003)]{gotz03}
Gotz D. et al. 2003, GCN 2459

\bibitem[Hawkins et al.(2003)]{hawkins03}
Hawkins E. et al. 2003, MNRAS, 346, 78


\bibitem[Lamb \& Reichart(2000)]{lamb00}
Lamb D.Q, \& Reichart D.E. 2000, ApJ, 536, 1

\bibitem[Linder \& Huterer(2003)]{linder02} 
Linder E.V. \& Huterer D. 2003, Phys. Rev. D, 67, 081303

\bibitem[Lloyd-Ronning, Fryer \& Ramirez-Ruiz(2002)]{lfr02} 
Lloyd--Ronning N.M., Fryer C.L. \& Ramirez--Ruiz E. 2002, 
ApJ, 574, 554

\bibitem[Nesseris \& Perivolaropoulos(2004)]{nesseris04} 
Nesseris S. \& Perivolaropoulos L. 2004, preprint (astro-ph/0401556)


\bibitem[Padmanabhan(2003)]{Padma03} 
Padmanabhan T. 2003, Phys. Reports, 380, 235

\bibitem[Peebles \& Ratra(2003)]{PR03}
Peebles P.J.E. \& Ratra, B. 2003, Rev.Mod.Phys., 75, 559

\bibitem[Perlmutter et al.(1999)]{perl99} 
Perlmutter S. et al. 1999, ApJ, 517, 565


\bibitem[Press et al.(1999)]{pre99} 
Press W.H. et al. 1999, {\it Numerical Recipes in C}, Cambridge University Press, 661

\bibitem[Riess et al. 1998]{riess98}
Riess A.G. et al. 1998, AJ, 116, 1009


\bibitem[Riess et al.(2004)]{riess04} 
Riess A.G. et al. 2004, ApJ, 607, 665

\bibitem[Rhoads (1997)]{r97} 
Rhoads J.E. 1997, ApJL, 487, L1

\bibitem[Sahni(2004)]{sahni03} 
Sahni V. et al. 2004, JETP Letters 77, 201, (astro-ph/0201498)

\bibitem[Sahni(2004)]{sahni04} 
Sahni V. 2004, Second Aegean Summer School 
on the Early Universe, in press (astro-ph/0403324)

\bibitem[Sari et al. (1999)]{sari99}
Sari R., Piran T. \& Halpern, J.P. 1999, ApJ, 519, L17

\bibitem[Schuecker et al. (2003)]{schu03}
Schuecker P. et al. 2003, A\&A, 402, 53

\bibitem[Schaefer(2003)]{schaefer}
Schaefer B.E. 2003, ApJ, 583, L67

\bibitem[Soderberg et al. 2004a]{sod04}
Soderberg A. et al. 2004a, Nature, 430, 648

\bibitem[Soderberg et al.  2004b]{sod04b} 
Soderberg A. et al. 2004b, ApJ, 606, 994



\bibitem[Visser (2004)]{visser04}
Visser M. 2004, preprint (gr-qc/0309109) 

\bibitem[Waxman 2004]{wax04}
Waxman E. 2004, ApJ, 605, L97

\bibitem[Weller \& Albrecht(2002)]{weller02} 
Weller J. \& Albrecht A. 2002, Phys. Rev. D,  65, 103512

\bibitem[Yamazaki et al.  2003]{yam03}
Yamazaki, R, Yonetoku D. \& Nakamura T. 2003, ApJ, 594, L79

\bibitem[Yonetoku et al.(2003)]{yonetoku04} 
Yonetoku D. et al. 2003, ApJ, 609, 935


\end{thebibliography}
\end{document}